\documentclass{article} 
\usepackage{iclr2019_conference,times}

\usepackage{amsmath,amsfonts,bm}

\newcommand{\newterm}[1]{{\bf #1}}

\def\eqref#1{equation~\ref{#1}}

\def\1{\bm{1}}

\def\veta{{\bm{{\eta}}}}

\def\vw{{\bm{w}}}
\def\vx{{\bm{x}}}

\DeclareMathAlphabet{\mathsfit}{\encodingdefault}{\sfdefault}{m}{sl}
\SetMathAlphabet{\mathsfit}{bold}{\encodingdefault}{\sfdefault}{bx}{n}

\def\sS{{\mathbb{S}}}

\newcommand{\pmodel}{p_{\rm{model}}}

\newcommand{\normmax}{L^\infty}

\DeclareMathOperator*{\argmax}{arg\,max}

\usepackage{hyperref}
\usepackage{graphicx}
\usepackage{url}
\usepackage{tabu}

\title{New CleverHans Feature: Better Adversarial Robustness Evaluations with Attack Bundling}

\author{Ian Goodfellow \thanks{\url{www.iangoodfellow.com}} \\
Google AI\\
\texttt{\{goodfellow\}@google.com} 
}

\iclrfinalcopy 
\begin{document}

\maketitle

\begin{abstract}
This technical report describes a new feature of the CleverHans \citep{cleverhans210}
  library called \newterm{attack bundling}.
Many papers about adversarial examples present lists of error rates
corresponding to different attack algorithms.
A common approach is to take the maximum across this list and compare
defenses against that error rate.
We argue that a better approach is to use attack bundling: the
max should be taken across many examples at the level of individual
examples, then the error rate should be calculated by averaging after
this maximization operation.
Reporting the bundled attacker error rate provides a lower bound on
the true worst-case error rate.
The traditional approach of reporting the maximum error rate across
attacks can underestimate the true worst-case error rate by an amount
approaching 100\% as the number of attacks approaches infinity.
Attack bundling can be used with different prioritization schemes
to optimize quantities such as error rate on adversarial examples,
perturbation size needed to cause misclassification, or failure rate
when using a specific confidence threshold.
\end{abstract}

\section{Introduction}

Many papers about adversarial examples present lists of error rates
corresponding to different attack algorithms.
A common approach is to take the maximum across this list and compare
defenses against that error rate.
We argue that a better approach is to use \newterm{attack bundling}: the
max should be taken across many examples at the level of individual
examples, then the error rate should be calculated by averaging after
this maximization operation.
Reporting the bundled attacker error rate provides a lower bound on
the true worst-case error rate.
The traditional approach of reporting the maximum error rate across
attacks can underestimate the true worst-case error rate by an amount
approaching 100\% as the number of attacks approaches infinity.
Attack bundling can be used with different prioritization schemes
to optimize quantities such as error rate on adversarial examples,
perturbation size needed to cause misclassification, or failure rate
when using a specific confidence threshold.
We contribute attack bundling to the {\tt cleverhans} \citep{cleverhans210}
library.

\section{Common reporting practices}

In papers about adversarial examples, it is common to see tables such as
Table \ref{table:mat}.
We call this a \newterm{Many Attack Table} (MAT).
MATs report the error rate when the
model is used to classify many different kinds of adversarial examples,
but there is no attempt to report the worst-case performance of the model.
MATs can be used to make some scientific points.
For example, the MAT can be used to show that the model performs well
against one attack and poorly against another attack.
If the model performs well against gradient-based attacks but poorly
against transfer-based attacks, a MAT could be used to argue
that the model suffers from gradient masking \citep{papernot2017practical}.
MATs are often used to argue that a specific defense
is strong because the defense performs well on many different kinds of
attacks. This is an incorrect use; if the defense performs badly on
even one kind of attack within the threat model under consideration,
then the model performs badly.
Attackers will choose the most effective attack, not a random attack
drawn from a long list of mostly ineffective attacks.

\begin{table}
  \centering
\begin{tabular}{c|c|c|c}
  No attack & Attack 1 & Attack 2 & Attack 3 \\
  \hline
  1\% & 3 \% & 11 \%  & 99\%
\end{tabular}
\caption{Error rates under different adversarial attacks.
  This table is a hypothetical illustration of a Many Attack Table (MAT)
  that should
  {\em } not be used to argue that a defense is strong.
This hypothetical table reports error rates for several different adversarial
attacks, rather than choosing the strongest attack against each example
then reporting the resulting error rate.
Showing multiple attacks without any report of the result of the maximization
operation can create the misleading impression that the defense performs
well because it performs well against several attacks. The existence of
one successful attack in fact shows that the defense has failed.
}
  \label{table:mat}
\end{table}

Another common type of table used in adversarial example papers is illustrated
in Table \ref{table:wat}. We call this a \newterm{Worst Attack Table} (WAT).
WATS are better than MATs: they attempt to describe the worst case by taking
the max across each row of a MAT. This models a scenario where the attacker
tries out many attack algorithms and then deploys the single attack algorithm
that causes the highest error rate. We argue that WATs are still not the correct
reporting format; they can underestimate the true error rate arbitrarily badly.

\begin{table}
  \centering
\begin{tabular}{c|c|c|c|c}
  No attack & Attack 1 & Attack 2 & Attack 3 & Max \\
  \hline
  1\% & 3 \% & 11 \%  & 99\%      &     {\bf 99} \%
\end{tabular}
\caption{Error rates under different adversarial attacks, and the maximum error rate
taken across different attacks in the table.
  This table is a hypothetical illustration of a Worst Attack Table (WAT)
  that we argue should
  {\em } not be used to argue that a defense is strong.
This WAT does a much better job of showing that that the proposed defense
  is {\em broken} than the MAT in Table \ref{table:mat} did.
Unfortunately, WATs can arbitrarily underestimate the true error rate.
WATs provide a lower bound on the true error rate but this bound may be very
loose, so they are useful for showing that defenses are broken but not for
showing that defenses work.
}
  \label{table:wat}
\end{table}

Unfortunately, WATs can badly underestimate the true error rate.
In general, attack algorithms can provide only a lower bound on the error rate
under a particular threat model: if you run an attack and the attack fools the
model, you know the model can be fooled in at least that instance, but if you
run an attack and the model is not fooled, you don't know whether or not there
exists a different attack that can fool the model for this example.
Unfortunately, attack algorithms remain the most popular method for evaluating
the robustness of models to adversarial examples. As long as this is the case,
it is important to design attack-based evaluation methodology to obtain the
tightest possible lower bound on the error rate.

\citet{max_confidence} introduced the concept of \newterm{attack bundling}.
In the attack bundling approach, may different attacks are run against each
example, just like when creating a MAT or a WAT.
The difference is that all of these attacks are then combined to create a single
stronger attack, in which the attacker chooses the best adversarial example
for each clean example.
Given a set of $n$ attack algorithms, the ``max'' column of a WAT produces
a tight lower bound on the true error rate if there exists one algorithm
that has optimal 0/1 loss for every example.
Attack bundling produces a bound that is at least as tight as the WAT bound.
The attack bundling bound is tight if for every example there exists one
algorithm that has optimal 0/1 loss for that example.
Table \ref{table:bundle_example} shows an example dataset and illustrates
how WAT can underestimate the error rate as well as how attack bundling
can obtain a tigher estimate of the error rate.

\begin{table}
  \centering
  \begin{tabular}{c|c|c|c}
    Example index & Error on Attack 1 & Error on Attack 2 & Error on Bundled Attacks \\
    \hline
    1             &  1                &     0             &  1                          \\
    2             &  0                &     1             &   1                        \\
    \hline
    Error rate    &  50\%             &     50\%          & 100\%                           \\
\end{tabular}
  \caption{An example situation where reporting the max error rate attacks
  underestimates the true worst-case error rate badly, but attack bundling
  finds the true error rate.
  In this example, we have a test set containing two examples,
  example $1$ in the first row and example $2$ in the second row.
  We have two attack algorithms, Attack 1 and Attack 2. In the second and third
  column where show an indicator variable describing whether the model makes
  an error on adversarial examples created by these algorithms.
  Attack 1 causes an error on example 1 and attack 2 causes an error on example 2,
  but the model gets the remaining adversarial examples correct.
  Now consider the bundled attack, where the attacker tries out both attacks
  and chooses the better one.
  As shown in the rightmost column, this causes an error for both examples.
  In the bottom row, we see the error rate for Attack 1 and Attack 2,
  as well as for the bundled attack.
  Because Attack 1 and Attack 2 both get an error rate of $50 \%$, a
  WAT would report an estimated worst-case error rate of $50 \%$.
  The bundled attack reveals that the worst-case error rate in this case
  is actually $100\%$.
  }
  \label{table:bundle_example}
\end{table}

In the worst case, a WAT can underestimate the true error rate by an amount
approaching 100\%.
Suppose that there is a dataset containing $n$ examples and there is a WAT
reporting the error rate for $n$ attacks. Suppose that attack $i$ results
in misclassification of example $i$ and correct classification of all other
examples.
The error rate for each attack is thus $\frac{1}{n}$ and the max error rate
across attacks is also $\frac{1}{n}$ but the error rate for attack bundling
is $1$.
Because $\lim_{n \rightarrow \infty} \frac{1}{n} = 0,$ the WAT can provide
an estimate of approximately $0\%$ error rate when the true worst-case error
rate is $100\%$.

While the WAT and MAT strategies involve reporting error rates in tables,
this is really only necessary if it is important to discuss differences
 in performance between attacks, for example to look for evidence of
 gradient masking.
 To measure the strength of a defense, it is not really necessary to report
 the error rate for the attacks. This can be summarized with just the
 results for the bundled attack.

\section{Discussion of threat models}

Attack bundling assumes that the attacker can try many attacks on one
example and see which is the most successful.

Under some black box threat models, this may not be the case. However, there
is not yet any clear methodology for how to evaluate defenses in the black
box setting.
\citet{athalye2018obfuscated} thus recommend evaluating defenses primarily
in the white box setting.
Even in the most limited black box settings, attackers could apply the
principle of fooling large ensembles of models \citep{liu2016delving}
to the bundling problem.
For example, the bundling algorithm could select the adversarial version
of each example that fools the most ensemble members, in hopes of increasing
the probability that the chosen adversarial exmaple will also fool the
completely unknown target model.

Some black box threat models do provide partial access to the model.
For example, the threat model may allow the attacker to send inputs to the
model and observe its output. This allows the attacker to try out all the
attacks in the bundle and then use the best one.

Some defenses are designed to interfere with the attacker in ways that could
interfere with attack bundling. For example, defenses based on using a
stochastic model make it impossible to know exactly which output will be
obtained when running a specific input. Trying $n$ attack points and then
choosing the best input to re-deploy may thus not perform as well as expected,
because a different output will be obtained on re-deployment.
Attack bundling can still be used in these cases, for example by choosing
the attack point that performs the best averaged across $m$ calls to the
stochastic model.
See \citet{carlini2017adversarial} for an example of how attackers can
circumvent stochastic defenses.

Some defenses try to discourage exploratory attacks by imposing negative
consequences for attackers who are detected.
For example, an account may be banned from a service if the account seems
to be trying out multiple attacks in order to bundle them.
Attackers in the white box setting can evade these kinds of defenses by
running the attacks on their own copy of the model, and then deploying
only the chosen attack against the target model.

The bundled attack also assumes that the attacker can try many attacks
before committing to one.
Real attackers may not be so powerful, even in a white box setting.
For example, if the attacker has a perfect copy of the model but the
starting point $\vx$ is not known ahead of time,
and the attacker has limited time to produce an adversarial example,
the attacker may not be able to bundle many attacks.

Many threat models in the adversarial example literature are based on an
attacker that can make norm-constrained perturbations of clean examples.
Attack bundling works in these threat models but also many other more
realistic threat models.
It is mostly agnostic to the action space of the attacker.

\section{Configuring an attack bundler}

Attack bundling algorithms can be configured with many different goals.
When an attacker attacks a batch of clean examples using a finite amount
of computation, the attacker must decide how to budget computation time
for each example.

An attacker who wants to maximize error rate should clearly stop spending
computation on an example after a misclassified adversarial example has
been found for that particular example.
Between two potential adversarial examples, the attacker prefers one that
results in misclassification.

For an attacker who wants to maximize failure rate \citep{max_confidence}
of a confidence thresholding defense, the attacker prefers to spend computation
on examples that do not yet exceed the confidence threshold.
An attacker who wants to maximize the failure rate \citep{max_confidence} of
a model that uses confidence thresholding as a defense faces some special
considerations.
When attacking an individual example, it is not necessary to know the defender's
specific threshold $t$ so long as $t \geq \frac{1}{2}$. When attacking a batch
of several examples, the attacker can obtain a better failure rate if the attacker
knows $t$ and spends more computation on examples for which the confidence of
incorrect predictions does not yet exceed $t$.
When choosing between two potential adversarial examples, the attacker prefers
the one with higher confidence.

Attackers can also use different example prioritization schemes and adversarial
selection schemes to achieve other goals, such as minimizing the perturbation
size needed for misclassification. Finding minimum-norm adversarial examples
can be useful because the examples can be sorted by perturbation norm and used
to make a plot with error rate on the vertical axis and the size of the norm
ball allowed by the threat model on the horizontal axis.

Attack bundling can also be used to re-implement existing adversarial example
strategies. For example, using $n$ random restarts of gradient-based optimization
\citep{madry2017towards} or $n$ random points in a random search procedure
\citep{athalye2018obfuscated}
can be regarded as bundling $n$ different attacks

\section{Example: attacking an MNIST model}

As an example, we show how attack bundling can be used to attack an MNIST model.
Here we used the regularized model of \citet{max_confidence}, which shows some
resistance to $\normmax$-constrained adversarial examples. We use a $\normmax$
constraint $\epsilon$ of $0.3$ for MNIST digits with values in $[0, 1]$.
We also clip the adversarial perturbations to this range.

\begin{figure}
  \centering
  \includegraphics[width=\textwidth]{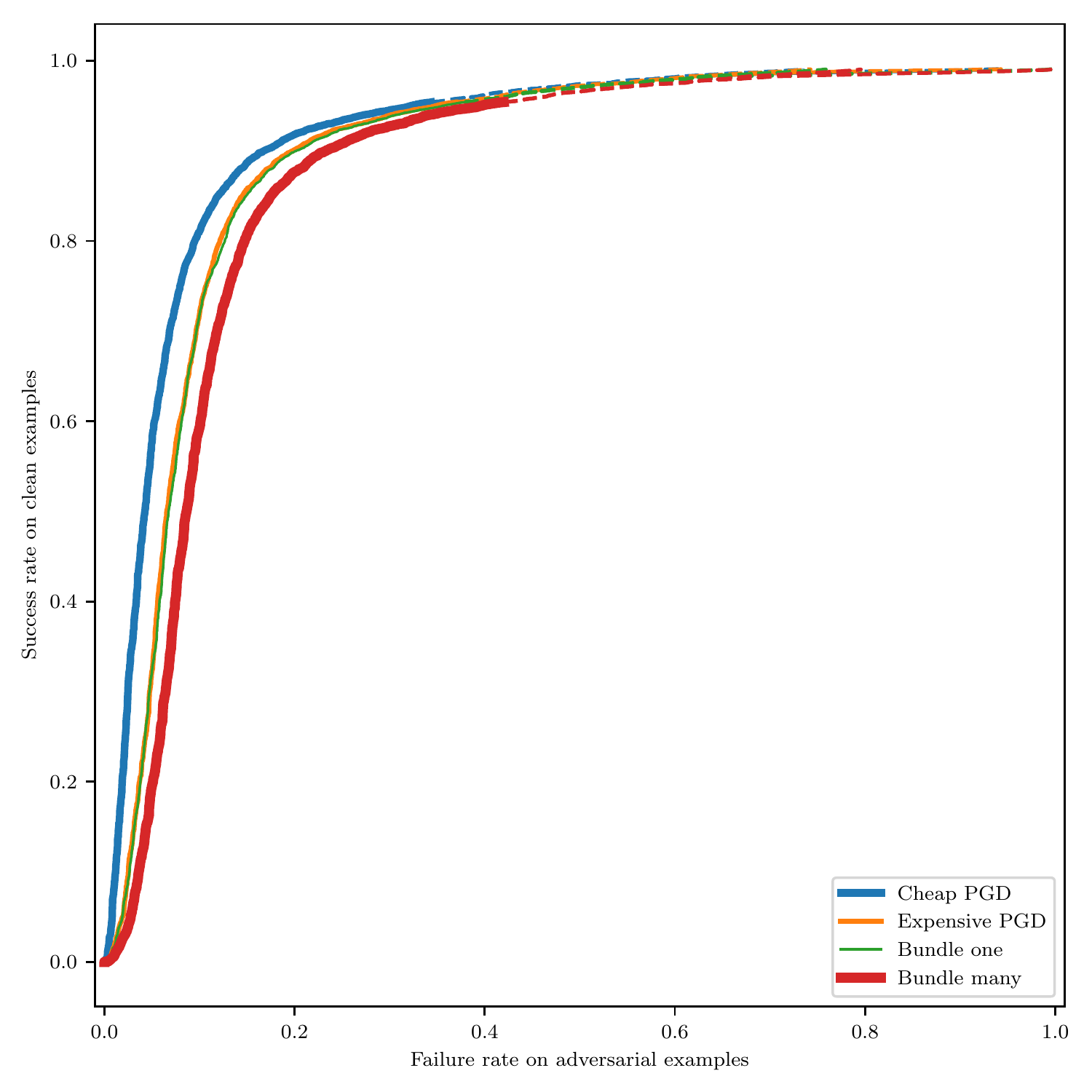}
  \caption{We compare the {\tt MaxConfidence} procedure \citep{max_confidence}
  using two different single attacks to the same procedure using bundled attacks.
  Both of the single attacks are randomly initialized projected gradient descent.
  We use two configurations of the attack: one that is relatively cheap using
  only $40$ steps with step size $0.1$, and another that is relatively expensive
  using $1000$ steps with step size $0.04$.
  The expensive attack is stronger than the cheap attack.
  However, a bundled attack using both of these and uniform noise attacks (to
  mitigate gradient masking) is just slightly stronger.
  Bundling multiple restarts rather than just a single start per attack results
  in greater strength.
  Rather than reporting each of these curves, and creating the psychological
  impression that the model performs as well as the upper-leftmost curve indicates,
  a paper evaluating this method as a defense should really report just the strongest
  bundled curve, since it has the tightest lower bound on the true failure rate.
  Reporting multiple curves in a separate graph may be useful for diagnosing
  gradient masking, etc., but is only distracting when the purpose is to argue
  that the defense is strong.
  See \citet{max_confidence} for more information about how to read these
  success-fail curves.
  }
\end{figure}

\section{Related work}

The general principle of choosing the strongest attack for each specific
context rather than evaluating each attack across all contexts is common
in computer security. We are not aware of the specific origin of this
idea.
\citet{athalye2018obfuscated} briefly described this principle in the context
of adversarial accuracy evaluations.
Our contributions are to: (1) provide a general, extendable implementation
in a standard library, (2) name the technique, (3) quantify the amount that
the alternative techniques can underestimate, (4) describe how bundling
algorithms apply to quantities other than accuracy, e.g. for making
success-fail curves, (5) describe how to apply bundling in the black box
setting, and (6) describe how bundling algorithms can prioritize
spending computation attacking different examples.

\section{Conclusion}

Attack bundling is now available as part of the {\tt cleverhans} library
\citep{cleverhans210}.
Papers evaluating defenses by running them against attacks should switch to the
new methodology.

\subsubsection*{Acknowledgments}

Many thanks to Nicholas Carlini for helpful discussions and to Nicolas Papernot
for serving as publication approval reviewer.

\bibliography{iclr2019_conference}

\begin{thebibliography}{7}
\providecommand{\natexlab}[1]{#1}
\providecommand{\url}[1]{\texttt{#1}}
\expandafter\ifx\csname urlstyle\endcsname\relax
  \providecommand{\doi}[1]{doi: #1}\else
  \providecommand{\doi}{doi: \begingroup \urlstyle{rm}\Url}\fi

\bibitem[Athalye et~al.(2018)Athalye, Carlini, and
  Wagner]{athalye2018obfuscated}
Anish Athalye, Nicholas Carlini, and David Wagner.
\newblock Obfuscated gradients give a false sense of security: Circumventing
  defenses to adversarial examples.
\newblock \emph{arXiv preprint arXiv:1802.00420}, 2018.

\bibitem[Carlini \& Wagner(2017)Carlini and Wagner]{carlini2017adversarial}
Nicholas Carlini and David Wagner.
\newblock Adversarial examples are not easily detected: Bypassing ten detection
  methods.
\newblock In \emph{Proceedings of the 10th ACM Workshop on Artificial
  Intelligence and Security}, pp.\  3--14. ACM, 2017.

\bibitem[Goodfellow et~al.(2018)Goodfellow, Qin, and Berthelot]{max_confidence}
Ian Goodfellow, Yao Qin, and David Berthelot.
\newblock Evaluation methodology for attacks against confidence thresholding
  models.
\newblock In \emph{Submitted to International Conference on Learning
  Representations 2019}, 2018.
\newblock URL \url{https://openreview.net/forum?id=H1g0piA9tQ}.
\newblock under review.

\bibitem[Liu et~al.(2016)Liu, Chen, Liu, and Song]{liu2016delving}
Yanpei Liu, Xinyun Chen, Chang Liu, and Dawn Song.
\newblock Delving into transferable adversarial examples and black-box attacks.
\newblock \emph{arXiv preprint arXiv:1611.02770}, 2016.

\bibitem[Madry et~al.(2017)Madry, Makelov, Schmidt, Tsipras, and
  Vladu]{madry2017towards}
Aleksander Madry, Aleksandar Makelov, Ludwig Schmidt, Dimitris Tsipras, and
  Adrian Vladu.
\newblock Towards deep learning models resistant to adversarial attacks.
\newblock \emph{arXiv preprint arXiv:1706.06083}, 2017.

\bibitem[Papernot et~al.(2017)Papernot, McDaniel, Goodfellow, Jha, Celik, and
  Swami]{papernot2017practical}
Nicolas Papernot, Patrick McDaniel, Ian Goodfellow, Somesh Jha, Z~Berkay Celik,
  and Ananthram Swami.
\newblock Practical black-box attacks against machine learning.
\newblock In \emph{Proceedings of the 2017 ACM on Asia Conference on Computer
  and Communications Security}, pp.\  506--519. ACM, 2017.

\bibitem[Papernot et~al.(2018)Papernot, Faghri, Carlini, Goodfellow, Feinman,
  Kurakin, Xie, Sharma, Brown, Roy, Matyasko, Behzadan, Hambardzumyan, Zhang,
  Juang, Li, Sheatsley, Garg, Uesato, Gierke, Dong, Berthelot, Hendricks,
  Rauber, Long, and McDaniel]{cleverhans210}
Nicolas Papernot, Fartash Faghri, Nicholas Carlini, Ian Goodfellow, Reuben
  Feinman, Alexey Kurakin, Cihang Xie, Yash Sharma, Tom Brown, Aurko Roy,
  Alexander Matyasko, Vahid Behzadan, Karen Hambardzumyan, Zhishuai Zhang,
  Yi-Lin Juang, Zhi Li, Ryan Sheatsley, Abhibhav Garg, Jonathan Uesato, Willi
  Gierke, Yinpeng Dong, David Berthelot, Paul Hendricks, Jonas Rauber, Rujun
  Long, and Patrick McDaniel.
\newblock Technical report on the cleverhans v2.1.0 adversarial examples
  library.
\newblock \emph{CoRR}, abs/1610.00768, 2018.
\newblock URL \url{http://arxiv.org/abs/1610.00768}.

\end{thebibliography}
\bibliographystyle{iclr2019_conference}

\appendix

\end{document}